\documentclass[aps,prl,reprint,groupedaddress]{revtex4-2}
\usepackage{graphicx}
\usepackage{amsmath}
\usepackage{amsfonts}
\usepackage{subfigure}
\begin{document}

%\preprint{}

\title{Universal Lower and Upper Bounds of Efficiency of Heat Engines \\ from Thermodynamic Uncertainty Relation}    
%Entropic Equality for the Precision of Current Fluctuation}


\author{Takaaki Monnai}
%\thanks{}
\affiliation{Department of Materials and Life Science, Seikei University, Tokyo, 180-8633, Japan}


\date{\today}

\begin{abstract}
According to Thermodynamics, the efficiency of a heat engine is upper bounded by Carnot efficiency. 
For macroscopic systems, the Carnot efficiency is, however, achieved only for quasi static processes. And, considerable attention has been paid to provide general evaluation of the efficiency at a finite speed. 
Recently, several upper bounds of the efficiency have been derived in the context of the trade-off among the efficiency, power, and other quantities such as the fluctuation of power. 

Here, we show universal lower and upper bounds of the efficiency from the thermodynamic uncertainty relations for the entropy production and for the heat transfers. The lower bound is characterized by the ratio between the fluctuation of the irreversible entropy production and mean output work. The upper bound of the efficiency is described by a generalized precision of the heat transfers among the working substance, and hot and cold reservoirs. We explicitly derive necessary and sufficient conditions of both the lower and upper bounds in a unified manner in terms of fluctuation theorem.     
Hence, our result provides an operating principle of the heat engine.
\end{abstract}

%\keywords{}

\maketitle

%\section{}
%\subsection{}
%\subsubsection{}
 {\it Introduction.---} 
According to thermodynamics, the efficiency of a heat engine $\eta$ given as the ratio between the input heat absorption and the output work is universally upper bounded by Carnot efficiency $\eta_C$. 
Since the Carnot efficiency is achieved only for quasi static processes for macroscopic systems, it is one of the most challenging problems to know the fundamental operating principle of the heat engines for general classical and quantum nonequilibrium systems including the molecular junction and nanowire\cite{Dubi1}.    
Therefore, considerable attention has been paid to some general trade-off relations between the efficiency and power.   
In particular, several explicit upper bounds of the efficiency at finite speed have been obtained beyond the linear response regime\cite{Tasaki1,Saito1,Pietzonka1,Goold1}.  
Furthermore, it has been pointed out that the fluctuation of the power is also a relevant quantity for the trade-off relation\cite{Pietzonka1,Goold1} for operation of steady state mesoscopic heat engines. 
Hence, the stochastic thermodynamics provides a comprehensive framework for this issue. 

On the other hand, the universal lower bound of the efficiency is still lacking. 
Such a lower bound of the efficiency is of fundamental interest for example to design a high-performance heat engine. 
In this Letter, as a first step toward this direction, we show universal lower and upper bounds of the efficiency $\eta$ of autonomous heat engines in terms of Carnot efficiency $\eta_C$ and the fluctuation of the entropy production caused by the  heat flowing among the working substance and reservoirs from thermodynamic uncertainty relations (TURs)\cite{Gingurich1,Gingurich2,Gingurich3,Seifert3}(see Ref.\cite{Horowitz1} for review) for the entropy production and heat transfers. 
In particular, the lower bound is expressed as  
\begin{eqnarray}
&&\eta\geq\frac{\eta_C}{1+k_BT_L{\rm Var}[\Sigma]/(2\langle W\rangle)}. \nonumber \\
%&\leq&\eta \nonumber \\
%&\leq&\frac{\eta_C}{1+2k_BT_L(\langle Q_1\rangle^2+\langle Q_2\rangle^2)/\left(({\rm Var}[Q_1]+{\rm Var}[Q_2])\langle W\rangle\right)}. \nonumber \\
&& \label{efficiency1}
\end{eqnarray}
The necessary and sufficient conditions of lower and upper bounds will be made explicit in terms of exchange fluctuation theorem (EFT)\cite{Jarzynski1,Timpanaro1,Monnai1}.     
Here, $k_B$ denotes Boltzmann constant, $T_L$ denotes the temperature of the cold reservoir, $k_B\Sigma$ is the entropy production, $W$ denotes the output work during a sufficiently long period $\tau$, $\langle\cdot\rangle$ and ${\rm Var}[\cdot]$ are the mean and variance. The dimensionless entropy production is equal to $\Sigma=-\beta_1 Q_1+\beta_2 Q_2$, where $\beta_1(\beta_2)$ denotes the inverse temperature of the hot(cold) reservoir, $Q_1$ and $Q_2$ stand for the net heat transfer from the hot reservoir $R_1$ to the working substance $S$ and that from $S$ to the cold reservoir $R_2$.    
Remarkably, the lower bound has a universal form fully characterized by the ratio between the fluctuation of the irreversible entropy production and the mean output work divided by $k_BT_L$.    
%We will also derive an upper bound (\ref{upper1}) from a generalization of TUR to multiple variables. %slightly extending the necessary and sufficient condition of TUR.
This lower bound also shows that the efficiency $\eta$ can be close to Carnot efficiency $\eta_C$ if the fluctuation of the entropy production is relatively small compared with the averaged work divided by $k_BT_L$. 

Note that the average of the work $\langle W\rangle$ and the variance of the entropy production ${\rm Var}[k_B\Sigma]$ are  typically proportional to the time duration $\tau$ in the stationary regime, and we can rewrite the lower bound as 
\begin{align}
&\frac{\eta_C}{1+k_BT_L{\rm Var}[\Sigma/\tau]\tau/(2\langle P\rangle)} \label{stationary1}
\end{align}
in terms of the entropy production per unit time and the power $P=W/\tau$.

Let us express the variance of the entropy production in terms of the heat transfer $Q_j$ ($j=1,2$), from which ${\rm Var}[\Sigma]$ can be experimentally accessible. For simplicity of notation, we consider the quantities during $\tau$ instead of those per unit time. The fluctuation of the entropy production in the denominator of (\ref{efficiency1}) can be expressed by the variance and covariance of the heat transfers of the heat engine
\begin{align}
&{\rm Var}[\Sigma]=\langle(-\beta_1Q_1+\beta_2Q_2)^2\rangle-\langle(-\beta_1Q_1+\beta_2Q_2)\rangle^2 \nonumber \\
%&=\beta_1^2{\rm Var}[Q_1]-2\beta_1\beta_2\langle (Q_1-\langle Q_1\rangle)(Q_2-\langle Q_2\rangle)\rangle+\beta_2^2{\rm Var}[Q_2] \nonumber \\
&=(-\beta_1,\beta_2){\rm Cov}[Q_1,Q_2](-\beta_1,\beta_2)^{\rm T} \nonumber \\ \label{covariance1}
\end{align} 
 and is usually considered to be the same order as the output work divided by $k_BT_L$, $\langle W\rangle/k_BT_L$. 
Eq. (\ref{covariance1}) plays an important role in deriving (\ref{minimum3}).
Here, ${\rm Cov}[\cdot,\cdot]$ and $(-\beta_1,\beta_2)^{\rm T}$ denote the variance-covariance matrix and a vector composed of the inverse temperatures of the reservoirs. 
Hence, our result (\ref{efficiency1}) provides a nontrivial practical lower bound.
Actually, the lower bound (\ref{efficiency1}) as well as an upper bound are numerically confirmed for a nanoscopic thermoelectric heat engine introduced in Ref. \cite{Esposito3} in the context of the efficiency at maximum power\cite{Curzon1,Seifert2}.   
\\

{\it Set up.---} For simplicity, let us consider the stationary set up, i.e., a working substance $S$ is simultaneously in contact with hot and cold reservoirs $R_1$ and $R_2$ at temperatures $T_H$ and $T_L$. %It is rather straightforward to consider the non stationary case, where the working substance $S$ is first connected only to the hot reservoir $R_1$, and detached from $R_1$ and subsequently connected to $R_2$.     
The heat transfer $Q_1(>0)$ from the hot reservoir $R_1$ to the working substance $S$ is partially transformed into the work $W(>0)$, and the heat $Q_2=Q_1-W(>0)$ is absorbed by the cold reservoir $R_C$. 

Before the initial time $t\leq 0$, the working substance $S$ is detached from the reservoirs. 
At the initial time $t=0$, the total system starts to interact, and after a sufficiently long period at $t=\tau$ the interactions to $R_1$ and $R_2$ are disconnected again. 

First, we derive the lower bound (\ref{efficiency1}) from the following thermodynamic uncertainty relation(TUR) for the entropy production
\begin{align}
&\frac{\langle\Sigma\rangle^2}{{\rm Var}[\Sigma]}\leq\frac{\langle\Sigma\rangle}{2}. \label{entropy1}
\end{align} 
Here, the standard TUR claims that the precision of some current-like quantities measured by the square mean to variance ratio (signal to noise ratio) is upper bounded by the half of the mean entropy production\cite{Seifert3,Gingurich1}. 

The mean entropy production is expressed in terms of $\eta$ as  
\begin{align}
&\langle\Sigma\rangle=\beta_2\langle W\rangle(\frac{\eta_C}{\eta}-1)%\leq \frac{1}{2}(-\beta_1,\beta_2)\Xi(-\beta_1,\beta_2)^{\rm T}
. \label{lower1}
\end{align}
Eq. (\ref{entropy1}) and (\ref{lower1}) provides the lower bound of the efficiency in (\ref{efficiency1}). 
And, the equality holds in (\ref{efficiency1}) if and only if TUR is saturated.  

Interestingly, a necessary condition, i.e, the lower bound (\ref{efficiency1}) should be smaller than the upper bound of the efficiency in terms of the fluctuation of power in Ref. \cite{Pietzonka1} is equivalent to TUR for $W$ being valid $\frac{\langle W\rangle^2}{{\rm Var}[W]}\leq\frac{\langle\Sigma\rangle}{2}$. 
This observation suggests that the lower bound of the efficiency (\ref{efficiency1}) is actually reasonable.    
\\

We can also derive an upper bound of the efficiency in terms of a multivariate precision of heat transfers 
\begin{align}
&\eta\leq\frac{\eta_C}{1+2k_BT_L\frac{\langle Q_1\rangle^2+\langle Q_2\rangle^2}{({\rm Var}[Q_1]+{\rm Var}[Q_2])\langle W\rangle}} \label{upper1}
\end{align}
from the following TUR for multiple variables  
\begin{align}
&(\langle Q_1\rangle,\langle Q_2\rangle)\Xi^{-1}(\langle Q_1\rangle,\langle Q_2\rangle)^{\rm T}\leq\frac{1}{2}\langle\Sigma\rangle. \label{multiple1}
\end{align} 
To complete the derivation of the upper bound under the necessary and sufficient condition of TUR for heat transfers (\ref{minimum4}), we show the following lemma. \\

\textbf{Lemma 1}\\
For an arbitrary invertible positive $M\times M$ symmetric matrix $V$ with $M\in\mathbb{N}$, the following inequality holds. 
For notational convenience, we use the bra-ket notation for convenience. ${}^\forall |{\bf a}\rangle=(a_1,a_2,...,a_M)^{\rm T}\in\mathbb{R}^M$, the $M\times M$ matrix $|{\bf a}\rangle\langle{\bf a}|$ satisfies 
\begin{align}
&|{\bf a}\rangle\langle{\bf a}|\leq\langle{\bf a}|V^{-1}|{\bf a})\rangle V. \label{inequality1}
\end{align}

The inequality for nonnegative matrices (\ref{inequality1}) is derived by noting that ${}^\forall\lambda\in\mathbb{R}$, ${}^\forall |{\bf c}\rangle\in\mathbb{R}^M$, the following inequality holds
\begin{align}
&(\lambda \langle{\bf a}|V^{-1}+\langle{\bf c}|)V(\lambda V^{-1}|{\bf a}\rangle+|{\bf c}\rangle)\geq 0. \label{positivity1}
\end{align}
The lemma 1 is immediately shown from the non-positivity of the discriminant of the function of $\lambda$ given by the left hand side of (\ref{positivity1}) for all $|{\bf c}\rangle$. 
 
Applying the lemma 1 to the vector $|{\bf a}\rangle=(\langle Q_1\rangle,\langle Q_2\rangle)^{\rm T}$ and the covariance matrix $V=\Xi$, it is straightforward to show 
the following lemma 2. \\

\textbf{Lemma 2}\\
The multivariate precision of the heat transfers and the covariance matrix $\Xi$ satisfy the following relation
\begin{align}
&\frac{\langle Q_1\rangle^2+\langle Q_2\rangle^2}{{\rm Var}[Q_1]+{\rm Var}[Q_2]}\leq(\langle Q_1\rangle,\langle Q_2\rangle)\Xi^{-1}(\langle Q_1\rangle,\langle Q_2\rangle)^{\rm T}. \label{inequality3}
\end{align}  
One can verify (\ref{inequality3}) by comparing the largest eigenvalues of both hand sides of (\ref{inequality1}).   

Combining (\ref{lower1}), (\ref{multiple1}) and Lemma 2, we obtain the upper bound (\ref{upper1}). 

{\it Microscopic foundation of the bounds.---}
The lower and upper bounds (\ref{efficiency1}) and (\ref{upper1}) are considered as universal from the generality of TUR. Actually, TUR holds for example for continuous time stochastic processes\cite{Gingurich1,Gingurich2,Gingurich3,Polettini1}, which are applied to the analysis of the efficiency of the molecular motors\cite{Pietzonka2} and of the microscopic heat engines\cite{Pietzonka1,Goold1}, and verified also for the overdamped Langevin dynamics\cite{Hasegawa2,Dechant1}. 
On the other hand, violations of TUR were reported for example for underdamped Langevin dynamics\cite{Hasegawa3} and general Hamiltonian dynamics\cite{Timpanaro1}. 

Under this circumstance, to explore the range of applicability of TURs and consequently of the lower and upper bounds (\ref{efficiency1}) and (\ref{upper1}), we show necessary and sufficient conditions of TURs for the entropy production and for the heat transfers in a unified manner in terms of the exchange fluctuation theorem(EFT)\cite{Jarzynski1,Timpanaro1} from a microscopic point of view. Such an analysis is useful. Actually, we show that the necessary and sufficient conditions of TURs for the entropy production (\ref{entropy1}) and for the heat transfers (\ref{multiple1}) are related but different.       
Here, the fluctuation theorem(FT) is a universal symmetry for the fluctuation of the entropy production, which relates the probabilities of the positive and negative entropy production\cite{Evans1,Gallavotti1,Kurchan1,Lebowitz1,Jarzynski2,Crooks1,Seifert1,Andrieux1} as reviewed in Ref. \cite{Esposito1,Seifert2}. FT was confirmed in various small systems such as a dragged colloidal particle in water\cite{Wang1}, nonequilibrium chemical reactions\cite{Gaspard1,Rao1}, electron transport in nanojunctions\cite{Utsumi1,Nakamura1} to name only a few.    
And, the mutual relation between FT and TUR has been intensively studied\cite{Timpanaro1,Monnai1,Pietzonka3,Pietzonka4,Hasegawa1}.       

First, we derive EFT for the heat transfer among the working substance and hot and cold reservoirs. 
Subsequently, we consider $N$ identical copies of this system, and define the rate function of the heat transfer of whole the system for finite $\tau$ owing to the large deviation principle. 
Then, we show necessary and sufficient conditions of the lower bound (\ref{efficiency1}) and the upper bound (\ref{upper1}) in a unified manner in terms of EFT. 

Without loss of generality, suppose the total system including the reservoirs obeys the Hamiltonian dynamics\cite{Jarzynski1}. 
The hot and cold reservoirs are initially prepared in local equilibrium states, which is described by the probability distribution $\rho(\Gamma)=\prod_{j=1,2}\frac{1}{Z_j}e^{-\beta_j H_j(\Gamma_j)}$. Here, $\Gamma$ and $\Gamma_j$ denote the state in the phase space of the total system and that of the $j$-th reservoir $R_j$. 
Also, $\beta_j$, $H_j(\Gamma_j)$, and $Z_j$ denote the inverse temperature, the Hamiltonian, and the partition function of the $j$-th reservoir, respectively.    

%In particular, as an essential step, we derive necessary and sufficient conditions of such TURs %by generalizing that of single variable case\cite{Monnai1} to multiple variables $Q_1$ and $Q_2$ 
%in terms of EFT. 
%Applying the EFT to the total system, we derive the lower bound from two-dimensional extension of TUR. 

Recall that the probability distribution of the heat transfer $P(Q_1,Q_2,\tau)$ satisfies EFT\cite{Jarzynski1} as  
\begin{align}
&P(Q_1,Q_2,\tau)=\int d\Gamma\rho(\Gamma)\delta(H_1(\Gamma_1)-H_1(\Gamma_1(\tau))-Q_1) \nonumber \\
&\;\;\;\;\;\;\;\;\;\;\;\;\;\;\;\;\;\;\;\;\;\;\; \times\delta(H_2(\Gamma_2(\tau))-H_2(\Gamma_2)-Q_2) \nonumber \\
&=\int d\Gamma\rho(\Gamma(\tau))e^{-\beta_1 Q_1+\beta_2 Q_2}\delta(H_1(\Gamma_1(\tau))-H_1(\Gamma_1)+Q_1) \nonumber \\
&\;\;\;\;\;\;\;\;\;\;\;\;\;\;\;\;\;\;\;\;\;\;\;\;\;\;\;\;\times\delta(H_2(\Gamma_2)-H_2(\Gamma_2(\tau))+Q_2) \nonumber \\
&=e^{-\beta_1 Q_1+\beta_2 Q_2}P(-Q_1,-Q_2,\tau), \label{EFT1}
\end{align}
where $\Gamma(\tau)$ and $\Gamma_j(\tau)$ denote the states of the total system and of the $j$-th reservoir at $t=\tau$, and $H_j(\Gamma_j(\tau))-H_j(\Gamma_j)$ is the heat absorbed by the reservoir $j$ during $\tau$. 
\\

{\it Exchange Fluctuation Theorem for Copies.---} 
To discuss the large deviation principle for finite $\tau$, let us consider $N$ identical and mutually independent copies of the system, i.e., the working substance $S$ and reservoirs $R_j$ $(j=1,2)$, and define the rate function in the limit  $N\rightarrow\infty$ as in Ref. \cite{Monnai1}. 
We identify the first copy as the original system. 
Since the heat transfer is additive, we define the net heat transfer of the $j$-th reservoir $Q_{tot,j}=\sum_{m=1}^N Q_j^{(m)}$. Here, $Q_j^{(m)}$ denotes the heat transfer of the $m$-th copy, which obeys the same statistics as $Q_j$ of the original system. 
The probability distribution of the net heat transfer of the total system also satisfies EFT
\begin{align}
&P_{tot}(Q_{tot,1}=NQ_1,Q_{tot,2}=NQ_2,\tau) \nonumber \\
&=e^{N(-\beta_1 Q_1+\beta_2 Q_2)}P_{tot}(Q_{tot,1}=-NQ_1,Q_{tot,2}=-NQ_2) \label{EFT2}
\end{align}
from the additivity of the heat transfers. 
Here, $P_{tot}(Q_{tot,1}=NQ_1,Q_{tot,2}=NQ_2,\tau)$ stands for the probability distribution that the net heat transfer $Q_{tot,j}$ of the ensemble of the reservoir $R_j$ equals to $NQ_j$ ($j=1,2$).  
The corresponding rate function 
\begin{align}
&I(Q_1,Q_2,\tau) \nonumber \\
&=-\lim_{N\rightarrow\infty}\frac{1}{N}\log P_{tot}(Q_{tot,1}=NQ_1,Q_{tot,2}=NQ_2,\tau) \nonumber \\
& \label{rate1}
\end{align}
is well-defined from the large deviation principle\cite{Ellis1}, since $Q_{tot,j}$ ($j=1,2$) is a sum of identical and independent stochastic variables.    
The rate function satisfies the following symmetry from EFT(\ref{EFT2})   
\begin{align}
&I(Q_1,Q_2,\tau)-I(-Q_1,-Q_2,\tau)=\beta_1 Q_1-\beta_2 Q_2. \label{EFT3}
\end{align} 

Let us evaluate the rate function from EFT (\ref{EFT3}) to provide a relation among the cummulants of the heat transfer  such as the mean, the variance, and the covariance. 
In the vicinity of the point corresponding to the mean value of the heat transfer ${\rm P_1}=(\langle Q_1\rangle,\langle Q_2\rangle)$, the rate function is well-approximated by the quadratic form
\begin{align}
&I_+(Q_1,Q_2,\tau) \nonumber \\
&=\frac{1}{2}(Q_1-\langle Q_1\rangle,Q_2-\langle Q_2\rangle)\Xi^{-1}(Q_1-\langle Q_1\rangle,Q_2-\langle Q_2\rangle)^{\rm T} \label{local1}
\end{align}
from the central limit theorem\cite{Feller1,Callen1}. 
Here, $\Xi$ is a short-hand notation of the $2\times 2$ variance-covariance matrix $\Xi={\rm Cov}(Q_1,Q_2)$ of the heat transfer.  
The subscript $+$ in $I_+$ emphasizes that (\ref{local1}) is a local approximation of the actual rate function.  
On the other hand, EFT (\ref{EFT3}) and (\ref{local1}) imply that the rate function behaves quadratically near another point ${\rm P}_2=(-\langle Q_1\rangle,-\langle Q_2\rangle)$ as
\begin{align}
&I_-(Q_1,Q_2,\tau) \nonumber \\
&=\frac{1}{2}(Q_1+\langle Q_1\rangle,Q_2+\langle Q_2\rangle)\Xi^{-1}(Q_1+\langle Q_1\rangle,Q_2+\langle Q_2\rangle)^{\rm T} \nonumber \\
&+\beta_1 Q_1-\beta_2 Q_2. \label{EFT4} 
\end{align}
\begin{figure}
\centering{
\subfigure[]{\includegraphics[scale=0.5]{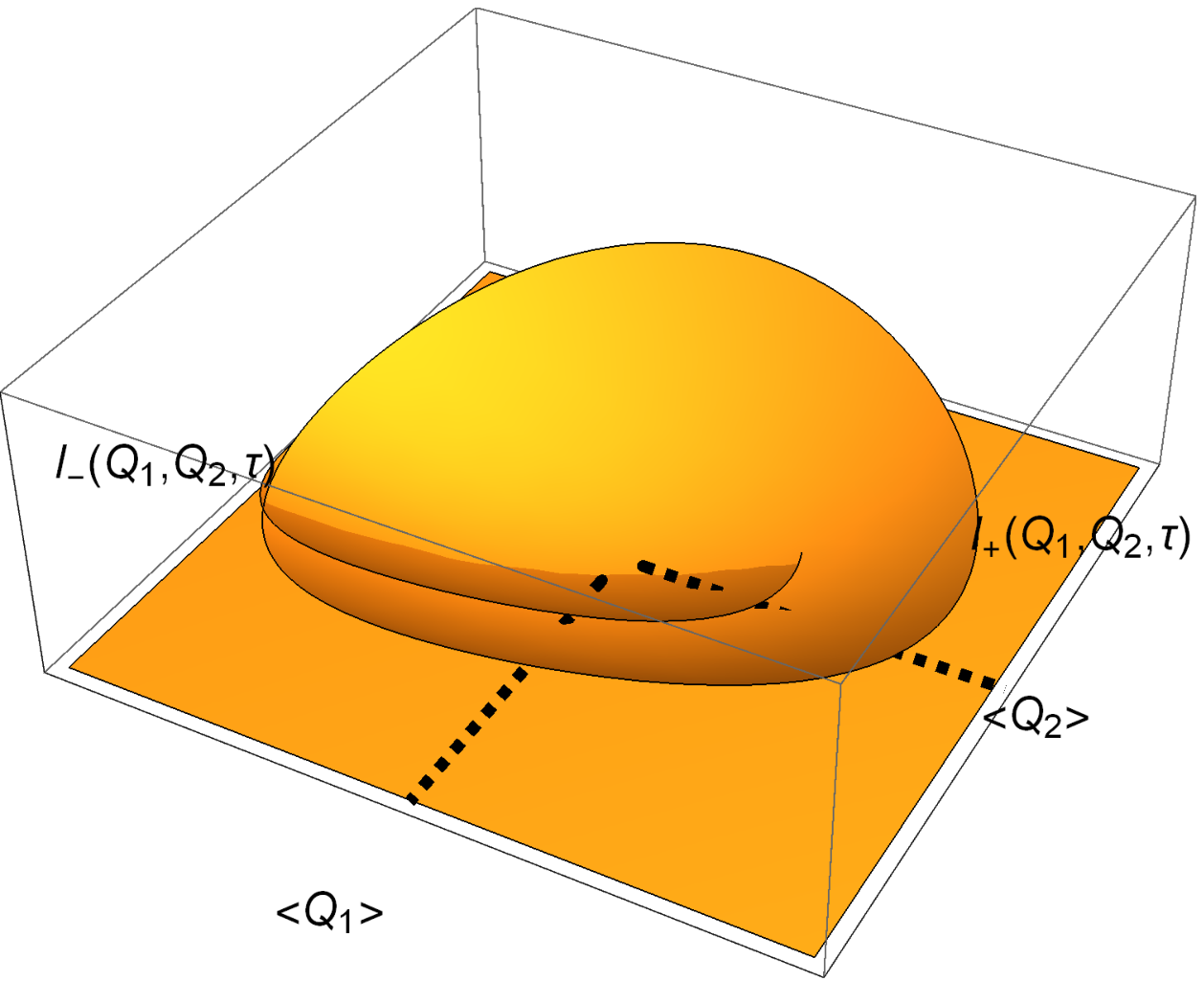}}\\
\subfigure[]{\includegraphics[scale=0.5]{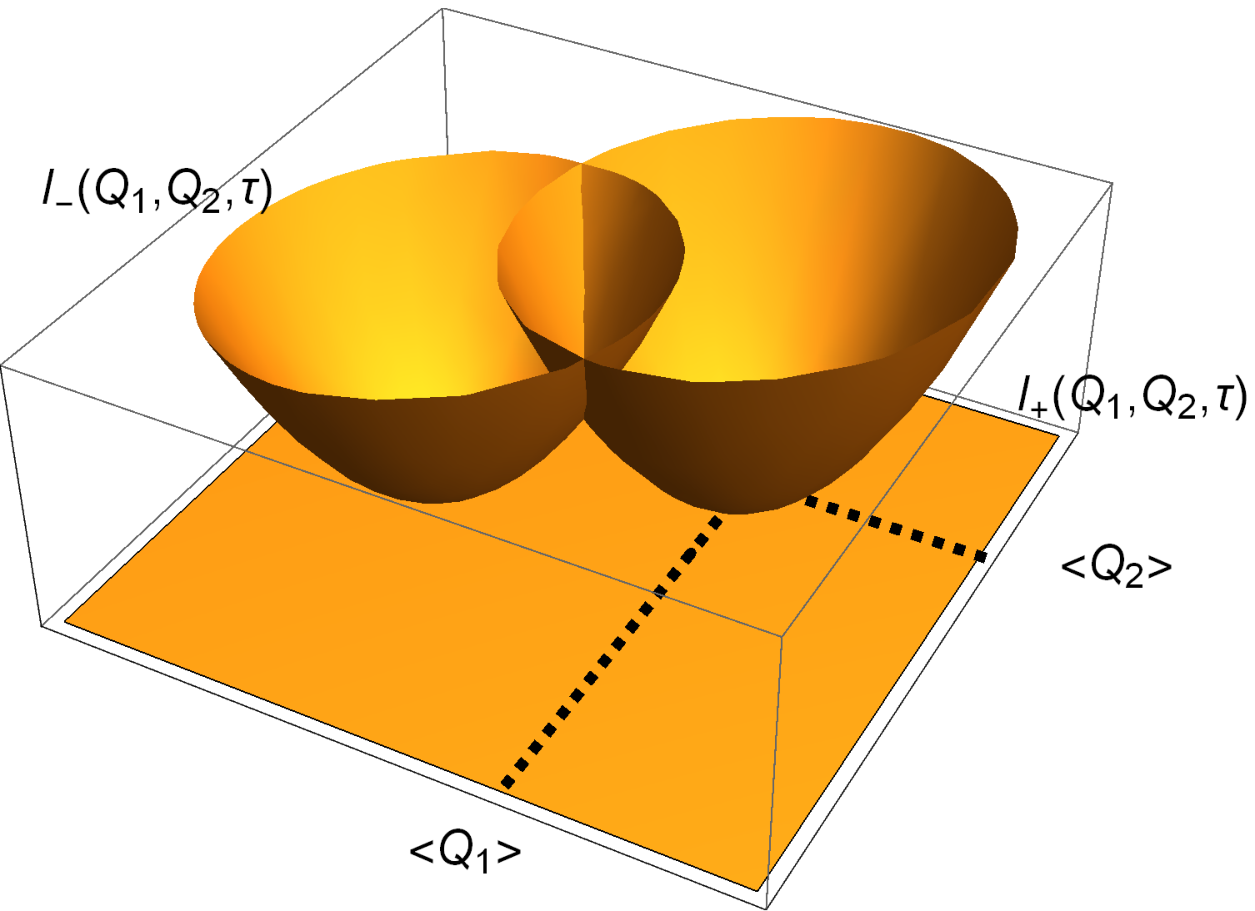}}%0-1.eps}
}
\caption{Schematic illustration of local approximations of the rate functions $I_+(Q_1,Q_2,\tau)$ and $I_-(Q_1,Q_2,\tau)$ located at ${\rm P}_1=(\langle Q_1\rangle,\langle Q_2\rangle)$ and ${\rm P}_2=(-\langle Q_1\rangle,-\langle Q_2\rangle,-\beta_1\langle Q_1\rangle+\beta_2\langle Q_2\rangle)$, respectively. (a)The case where TUR holds. The surface corresponding to the local approximation $I_-(Q_1,Q_2,\tau)$ crosses the $Q_1-Q_2$ plane from the necessary and sufficient condition of TUR. (b)The case of TUR being violated. The curvatures of the local approximations at ${\rm P}_1$ and ${\rm P}_2$ are large so that the $I_-(Q_1,Q_2,\tau)$ does not cross the $Q_1-Q_2$ plane.}   
\end{figure}

{\it Lower bound of the Efficiency.---} 
First, we show a necessary and sufficient condition of TUR for the entropy production (\ref{entropy1}) in terms of EFT. 
\\

\textbf{Necessary and Sufficient Condition of TUR}\\ 
TUR (\ref{entropy1}) holds if and only if the minimum of the local approximation $I_-(Q_1,Q_2,\tau)$ is non-positive.    
\\

To derive this necessary and sufficient condition of TUR, let us investigate the condition for the minimum of $I_-(Q_1,Q_2)$ being non-negative
\begin{align}
&{\rm min}_{Q_1,Q_2}I_-(Q_1,Q_2)\leq 0. \label{minimum1}
\end{align} 
By taking the partial derivatives of $I_-(Q_1,Q_2)$ with respect to $Q_1$ and $Q_2$, and equating zero, one can easily verify that the minimum is achieved for 
\begin{align}
&(Q_1+\langle Q_1\rangle,Q_2+\langle Q_2\rangle)^{\rm T}=\Xi(-\beta_1,\beta_2)^{\rm T}. \label{minimum2}
\end{align}
Substituting (\ref{minimum2}) into (\ref{EFT4}) and applying (\ref{covariance1}), the minimum of the local approximation $I_-(Q_1,Q_2)$ is rewritten as  
\begin{align}
&{\rm min}_{Q_1,Q_2}I_-(Q_1,Q_2)=-\frac{1}{2}{\rm Var}[\Sigma]+\langle\Sigma\rangle. \label{minimum3}
\end{align}
Hence, (\ref{minimum1}) is equivalent to TUR (\ref{entropy1}). 
This completes the proof. \\

{\it Upper bound of the Efficiency.---}
Remarkably, the necessary and sufficient condition of (\ref{multiple1}) is different from (\ref{minimum1}) 
\begin{align} 
&I_-(\langle Q_1\rangle,\langle Q_2\rangle)\leq 0. \label{minimum4}
\end{align}
Note that (\ref{minimum4}) is a sufficient condition of TUR for the entropy production (\ref{entropy1}) , since if (\ref{minimum4}) holds then ${\rm min}_{Q_1,Q_2}I_-(Q_1,Q_2)\leq 0$ is valid. %negative at $(Q_1,Q_2)=(\langle Q_1\rangle, \langle Q_2\rangle)$ then 
We also remark that (\ref{minimum4}) is reasonable, since the non-positivity of the local approximation at the mean value is actually equivalent to TUR for single variable case\cite{Monnai1}.    
\\

{\it Example.---}
\begin{figure}
\center
{\subfigure[]{\includegraphics[scale=0.5]{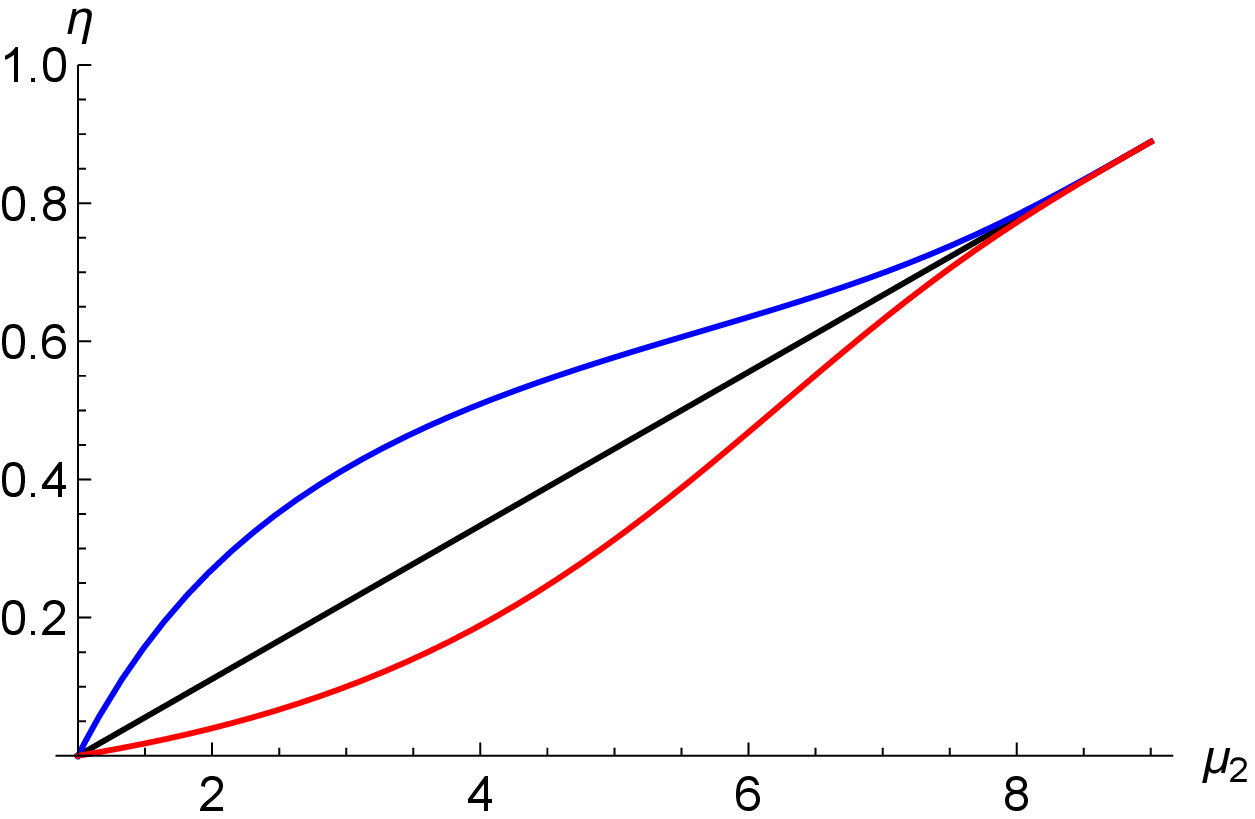}}
\subfigure[]{\includegraphics[scale=0.5]{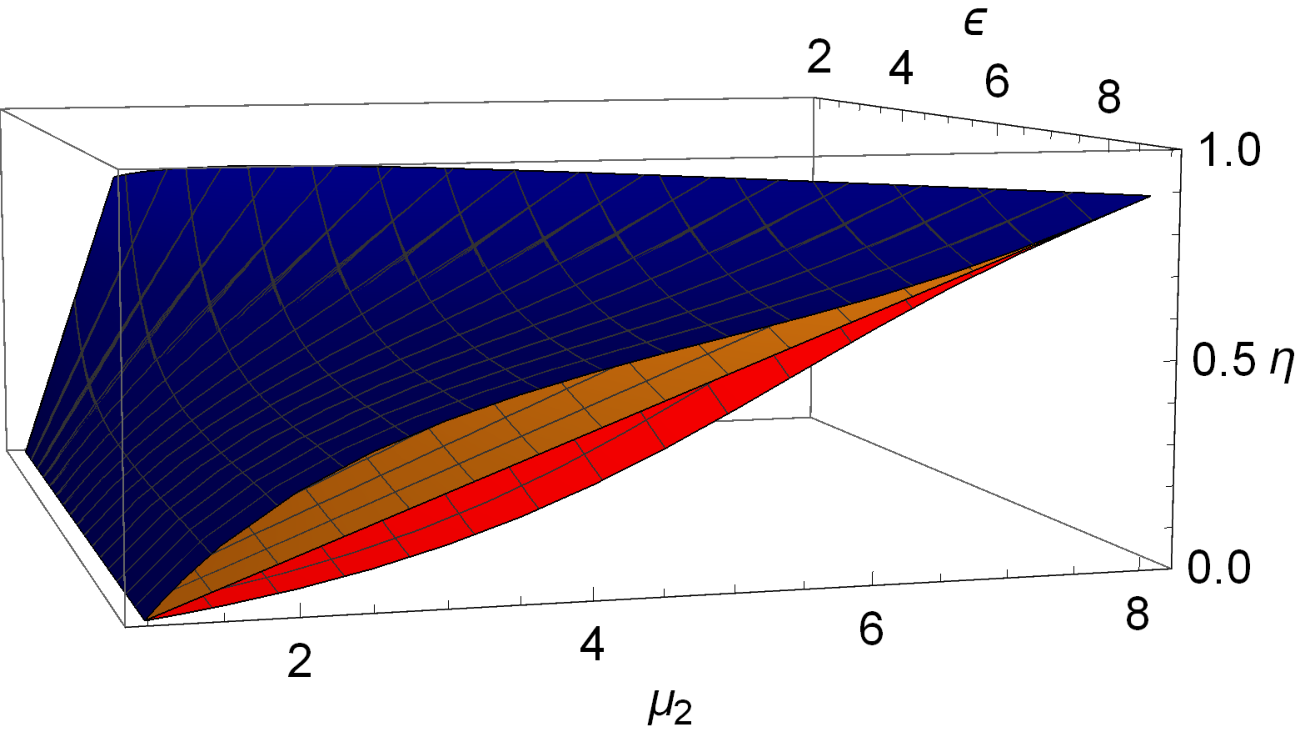}}%{efficiencyJan10.eps}}
} 
\caption{The lower- and the upper-bound of the efficiency (\ref{efficiency1}) and (\ref{upper1}) for a quantum thermoelectric heat engine. (a)The efficiency $\eta$ (black curve) and its lower and upper bounds (red and blue curves) as a function of $\mu_1$ for $\Gamma_1=\Gamma_2=1$, $\epsilon=10$, $\mu_2=1$, $T_1=10$, and $T_2=1$.   
(b)The efficiency $\eta$ (orange) and its lower and upper bounds (red and blue surfaces) as a function of $\mu_1$ and $\epsilon$. The parameters except for $\epsilon$ is the same as in (a).}
\end{figure}
Let us confirm the validity of (\ref{efficiency1}) for the nanothermoelectric heat engine introduced in Ref. \cite{Esposito3}. 
The electron flows among two reservoirs through a quantum dot with a sharply defined resonant energy $\epsilon$.  
The temperature $T_i$ and the chemical potential $\mu_i$ ($i=1,2$) satisfy $T_1>T_2$ and $\mu_1<\mu_2$, and the system operate as a heat engine with the input heat $\epsilon-\mu_1$ and output work $\mu_2-\mu_1$ per an electron transfer by choosing the thermodynamic affinity $-\frac{\epsilon-\mu_1}{T_1}+\frac{\epsilon-\mu_2}{T_2}$ positive. 
The time evolution of this heat engine is described by a stochastic master equation for the occupation number of the dot.  The transition of the electron in the dot to the reservoir $R_j$ occurs at the rate $k_{01}^{(j)}=\Gamma_j f(x_j)$ and the corresponding reversed transition from the reservoir to the empty dot occurs with the rate $k_{10}^{(j)}=\Gamma_j(1-f(x_j))$ by using the wideband approximation. Here, we used abbreviated notation $x_j=\frac{\epsilon-\mu_j}{T_j}$ and $f(x)=\frac{1}{e^{x}+1}$ denotes the Fermi-Dirac distribution. $\Gamma_j$ stands for the frequency. 
And, the cumulants of the number of transfered particles can be calculated in the general method in terms of the transition rates\cite{Baisei1}. 
The number of transfered particles from the reservoir $R_1$ to the dot $J_1$ and that from the dot to the clod reservoir $J_2$ are related to the heat transfers as $Q_1=(\epsilon-\mu_1)J_1$ and $Q_2=(\epsilon-\mu_2)J_2$, and the means  
and the variances of the accumulated number of transfered particles are given as 
\begin{align}
&\langle J_1\rangle =\langle J_2\rangle =\frac{\Gamma_1\Gamma_2}{\Gamma_1+\Gamma_2}\left(f(x_1)-f(x_2)\right)\tau \label{mean1}
\end{align}
 and 
\begin{align}
&{\rm Var}[J_1]={\rm Var}[J_2]={\rm Cov}[J_1,J_2] \nonumber \\
&=(\frac{\Gamma_1\Gamma_2}{\Gamma_1+\Gamma_2}\left(f(x_1)(1-f(x_2)\right) \nonumber \\
&+f(x_2)(1-f(x_1)))-\frac{1}{\Gamma_1+\Gamma_2}\langle J_1\rangle^2)\tau. \label{variance1}
\end{align} 
The efficiency is calculated as 
\begin{align}
&\eta=\frac{\mu_2-\mu_1}{\epsilon-\mu_1}. \label{efficiency2}
\end{align}
In this manner, the applicability of lower and upper bounds (\ref{efficiency1}) and (\ref{upper1}) are verified as illustrated in Fig. 2.

{\it Conclusion.---}\\
We have shown universal lower and upper bounds of the efficiency of a class of mesoscopic heat engines from TURs. As for the lower bound, the ratio between the variance of the dimensionless entropy production due to the heat transfer $\Sigma$ and the mean of the power $\beta_2\langle P\rangle$ plays a role. 
On the other hand, the upper bound holds under a necessary and sufficient condition of multivariate TUR for heat transfers explicitly defined in terms of the fluctuation of the heat transfer. 
\begin{acknowledgments}
This work was supported by the Grant-in-Aid for Scientific Research (C) (No.~18K03467 and No.~22K03456) from the Japan Society for the Promotion of Science (JSPS).
\end{acknowledgments}

\end{document}